\documentstyle[preprint,aps]{revtex}
\renewcommand{\mathbf}[1]{{\bf#1}}
  
\begin{document}
\draft
\title{SEMICLASSICAL EXPLANATION OF THE MATTEUCCI-POZZI AND AHARONOV-BOHM PHASE SHIFTS}
\author{Timothy H. Boyer}
\address{Department of Physics, City College of the City University of New York, New York, NY 10031}
\date{\today}
\maketitle

\begin{abstract}
Classical electromagnetic forces can account for the experimentally observed phase shifts seen in an electron interference pattern when a line of electric dipoles or a line of magnetic dipoles (a solenoid) is placed between the electron beams forming the interference pattern.  
\end{abstract}

\noindent{\bf INTRODUCTION}

     When a line of dipoles is placed between the beams of electrons forming an interference pattern, there is a shift in the interference pattern.  In the case of a line of electric dipoles, the interference pattern shift was observed experimentally by Matteucci and Pozzi.[1]  In the case of a line of magnetic dipoles (a solenoid), the phase shift was suggested by Aharonov and Bohm[2] and was first observed experimentally by Chambers[3].  Both these phase shifts take the same experimental form and both allow semiclassical interpretations as arising from lag effects due to classical electromagnetic forces.  Only the second phase shift is mentioned in the textbooks[4] because it is purported to occur in the absence of classical forces.  Here we will sketch the parallels between the electric and magnetic phase shifts and then discuss the forces in the magnetic case.

\noindent{\bf THE MATTEUCCI-POZZI PHASE SHIFT} 

     The experimental system used by Matteucci and Pozzi[1] may be approximated by a beam of charged particles which passes through two slits in a partition and produces a particle interference pattern on a screen.  (See Ref. 8, Fig. 3.)  In the region to the right beyond the slits, charges move in the $y$-direction with velocity $\mathbf{v}_0=\mathbf{\hat{j}}v_0$ along the lines $x=\pm d$, $y=v_0t$, $z=0$ with the choice of +d or -d referring to the two slits.

     The experiment of Mattucci and Pozzi involves the insertion of a line of electric dipoles between the two charged-particle beams.  Thus imagine two line charges $+\lambda$ and $-\lambda$ extending parallel to the $z$-axis and placed at $y=0$, $x=+\epsilon$, and $y=0$, $x=-\epsilon$, where the distance $\epsilon$ is small compared to the separation $2d$ between the beams, $\epsilon<<d$.  The line charges separated by $2\epsilon$ produce a line of electric dipoles with dipole moment per unit length $\mathbf{\hat{i}}\wp=\mathbf{\hat{i}}2\epsilon\lambda$.  

     A charge $e$ in the particle beam at coordinates $x_e$, $y_e$, $z_e=0$ has an electrostatic field
\begin{equation}
\mathbf{E}_e(x,y,z)=\frac{e[\mathbf{\hat{i}}(x-x_e)+\mathbf{\hat{j}}(y-y_e)+\mathbf{\hat{k}}(z)]}{[(x-x_e)^2+(y-y_e)^2+z^2]^{3/2}}
\end{equation}
and so applies an electrostatic force on the line of dipoles[5]
\begin{equation}
\mathbf{F}_{\wp}=\int^{\infty}_{-\infty}dz(\wp\mathbf{\hat{i}}\cdot\nabla)\mathbf{E}_e=\int dz\wp\left(\frac{\partial}{\partial x}\mathbf{E}_e\right)_{x=0,y=0}
\end{equation}
The component of force in the y-direction is
\begin{equation}
F_{\wp y}=\int^{\infty}_{-\infty}dz\left( \frac{e\wp 3(-y_e)(+x_e)}{(x_e^2+y_e^2+z^2)^{5/2}}\right)=-e\wp\frac{4x_ey_e}{(x_e^2+y_e^2)^2}
\end{equation}
Since electrostatic forces satisfy Newton's third law, the line of dipoles causes a force $\mathbf{F}_e=-\mathbf{F}_\wp$ on the passing charge.

     Now a charge $e$ which passes through the slit on the side $x=+d$ is closer to the $+\lambda$ line charge and is repelled, and thus slows down as it approaches the line of dipoles; after passing the line of dipoles, the repulsion causes the charge to speed up.  A charge passing on the opposite side $x=-d$ is attracted to the closer negative line charge and so first speeds up and then, after passing the line of dipoles, is slowed down.  Thus two charges which leave the source traveling side-by-side will have a relative lag after they pass the line of dipoles.

     Compared to a charge which moves with constant velocity $\mathbf{v}_0=\mathbf{\hat{j}}v_0$, the change in the y-component of velocity $\Delta v_y^{(+)}(t)$ of the charge traveling on the side $x=+d$ is given by
\begin{equation}
\Delta v^{(+)}_y(t)=\frac{1}{m}\int^{t'=t}_{-\infty}dt'F_{\wp y}(t')\simeq\frac{e\wp}{m}\int_{-\infty}^{y'=y_e}\frac{dy'}{v_0}\frac{4x_ey'}{(x_e^2+y'^2)^2}=-\frac{e\wp}{mv_0}\frac{2x_e}{(x_e^2+y_e^2)}
\end{equation}
For a charge which has passed the line of dipoles, this change in velocity leads to a relative displacement compared to a charge moving with constant velocity
\begin{equation}
\Delta y^{(+)}=\int^\infty_{-\infty}dt\Delta v^{(+)}_y(t)\simeq\int_{-\infty}^{\infty}\frac{dy}{v_0}\left(\frac{-e\wp}{mv_0}\frac{2x_e}{(x_e^2+y^2)}  \right)=-\frac{2\pi e\wp}{mv_0^2}
\end{equation}
The relative displacement $\Delta y^{(-)}$ for a charge traveling on the side $x=-d$ is reversed in sign.  It follows that the relative displacement between charges passing on opposite sides of the line of dipoles is
\begin{equation} 
\Delta Y=\Delta y^{(+)}-\Delta y^{(-)}=-\frac{4\pi e\wp}{mv_0^2}
\end{equation}
We note that the relative displacement does not depend upon the magnitude of the separation $d$ between the passing charges and the line of dipoles.  Simply passing the line of dipoles produces the lag $\Delta y^{(+)}$ or $\Delta y^{(-)}$, and the resulting relative lag $\Delta Y$.

     Thus far our calculation has been entirely classical.  At this point we wish to make contact with the quantum interference pattern produced by the charges.  The relative displacement $\Delta Y$ between particles of momentum $\mathbf{p}=m\mathbf{v}_0$ passing on opposited sides of the line of electric dipoles $\mathbf{\wp}$ corresponds to a relative phase shift $\Delta\phi_\wp$ between the particle wave functions
\begin{equation}
\Delta \phi_{\wp}=\frac{p_y\Delta Y}{\hbar}=-\frac{mv_0}{\hbar}\frac{4\pi e\wp}{mv_0^2}=-\frac{4\pi e\wp}{v_0\hbar}=-\frac{4\pi e(2\epsilon\lambda)}{v_0\hbar}
\end{equation} 
The relative phase shift is complensated by a deflection of the double-slit interference pattern arising from the two beams.  Just this phase-shift deflection was observed by Matteucci and Pozzi in 1985[1].

\noindent{\bf THE AHARONOV-BOHM PHASE SHIFT}

     The Aharonov-Bohm phase shift can be pictured with  the same electron beam arrangement, but now instead of a line of electric dipoles, we insert rather a line of magnetic dipoles between the beams.  A line of magnetic dipoles, which can be pictured as a stack of current loops, is just a solenoid.  Here we place a long, thin solenoid of cross-sectional area $\sl{A}$ and interior magnetic field $B_0$ so that its axis of symmetry is along the $z$-axis of coordinates.  Then the azymuthal surface currents per unit length are given by $\mathbf{K}=\mathbf{\hat{\phi}}B_0c/4\pi$ and the magnetic dipole moment per unit length is 
$\mathbf{\hat{k}}\mu=\mathbf{\hat{k}}KA/c=\mathbf{\hat{k}}B_0A/4\pi$.

     A charged particle located at $x_e$, $y_e$, $z_e=0$ and moving with velocity $\mathbf{v}_0=\mathbf{\hat{j}}v_0$ causes a magnetic field
\begin{equation}
\mathbf{B}_e(x,y,z)=\frac{ev_0}{c}\frac{[\mathbf{\hat{i}}z-\mathbf{\hat{k}}(x-x_e)]}{[(x-x_e)^2+(y-y_e)^2+z^2]^{3/2}}
\end{equation}
and so puts a net force on the solenoid[6]
\begin{equation}
\mathbf{F}_\mu=\int dz\nabla(\mu\mathbf{\hat{k}}\cdot\mathbf{B}_e)=\int dz\left[\left(\mathbf{\hat{i}}\frac{\partial}{\partial x}+\mathbf{\hat{j}}\frac{\partial}{\partial y}+\mathbf{\hat{k}}\frac{\partial}{\partial z}\right)\frac{\mu ev_0}{c}\frac{-(x-x_e)}{[(x-x_e)^2+(y-y_e)^2+z^2]^{3/2}}  \right]_{x=0,y=0}
\end{equation}
The component of force in the y-direction is
\begin{equation}
F_{\mu y}=\int_{-\infty}^{\infty}dz\left(\frac{\mu ev_03(+x_e)(+y_e)}{c(x_e^2+y_e^2+z^2)^{5/2}} \right)=+\frac{e\mu v_0}{c}\frac{4x_ey_e}{(x_e^2+y_e^2)^2}
\end{equation}
We notice that this force $F_{\mu y}$ has exactly the same functional form as was found in Eq.(3) for $F_{\wp y}$ in the electrostatic case above.

     If we assume that the net force between the solenoid and the passing charge satisfies Newton's third law, then there is a force $\mathbf{F}_e$ on the passing charge, $\mathbf{F}_e=-\mathbf{F}_\mu $, and, just as in the electric dipole case, there is a relative lag effect for charges passing on opposite sides of the solenoid.  Exactly as in the electric case, we can calculated the relative change in velocity
\begin{equation}
\Delta v^{(+)}_y(t)=\frac{1}{m}\int^{t'=t}_{-\infty}dt'F_{e y}(t')\simeq-\frac{e\mu v_0}{mc}\int_{-\infty}^{y'=y_e}\frac{dy'}{v_0}\frac{4x_ey'}{(x_e^2+y'^2)^2}=+\frac{e\mu }{mc}\frac{2x_e}{(x_e^2+y_e^2)}
\end{equation}
the relative displacements
\begin{equation}
\Delta y^{(+)}=\int^\infty_{-\infty}dt\Delta v^{(+)}_y(t)\simeq\int_{-\infty}^{\infty}\frac{dy}{v_0}\left(\frac{+e\mu}{mc}\frac{2x_e}{(x_e^2+y_e^2)}  \right)=+\frac{2\pi e\mu}{mv_0c}
\end{equation}
\begin{equation} 
\Delta Y=\Delta y^{(+)}-\Delta y^{(-)}=+\frac{4\pi e\mu}{mv_0c}
\end{equation}
and the semiclassical phase shift due to the magnetic dipoles 
\begin{equation}
\Delta \phi_{\mu}=\frac{p_y\Delta Y}{\hbar}=\frac{mv_0}{\hbar}\frac{4\pi e\mu}{mv_0c}=\frac{4\pi e\mu}{c\hbar}=\frac{eB_0\sl{A}}{c\hbar}
\end{equation} 
This phase shift was first observed experimentally by Chambers and was subsequently confirmed by numerous researchers[3].

\noindent{\bf AGREEMENT WITH SCHROEDINGER EQUATION}

     The semiclassical explanations given here for the Matteucci-Pozzi and Aharonov-Bohm phase shifts are in agreement with the application of the WKB approximation to the nonrelativistic Schroedinger equation.

     In the electrostatic case, the charges pass the line of electric dipoles with an electrostatic potential 
\begin{equation}
\Phi_\lambda(\mathbf{r})=-2\lambda\ln[(x-\epsilon)^2+y^2]^{1/2}+2\lambda\ln[(x+\epsilon)^2+y^2]^{1/2}=\frac{2\wp x}{x^2+y^2}
\end{equation}
with $\wp=2\epsilon\lambda$ giving a particle Hamiltonian
\begin{equation}
H=\frac{p^2}{2m}+e\Phi_\lambda
\end{equation}
The shift in the double slit interfence pattern can be calculated[7] by evaluating the phase along the two different paths on opposite side of the line of dipoles
\begin{eqnarray*}
\Delta \phi_\wp=\frac{1}{\hbar}\int dy(p_y)_{x=+d}-\int dy(p_y)_{x=-d}=\frac{1}{\hbar}\int dy(p_0^2-2me\Phi_\lambda)^{1/2}_{x=+d}-\int dy(p_0^2-2me\Phi_{\lambda})^{1/2}_{x=-d}
\end{eqnarray*}
\begin{eqnarray*}
\simeq\frac{1}{\hbar}\int dy\left[p_0\left(1-\frac{me\Phi_\lambda}{p_0^2}\right)\right]_{x=+d}-\frac{1}{\hbar}\int dy\left[p_0\left(1-\frac{me\Phi_\lambda}{p_0^2}\right)\right]_{x=-d}
\end{eqnarray*}
\begin{equation}
=\frac{1}{\hbar }\int dy\left(\frac{-2me\wp x}{p_0(x^2+y^2)}\right)_{x=+d}-\frac{1}{\hbar}\int dy\left(\frac{-2me\wp x}{p_0(x^2+y^2)}\right)_{x=-d}
=-\frac{4\pi e\wp}{v_0\hbar}=-\frac{4\pi e(2\epsilon\lambda)}{v_0\hbar}
\end{equation}
This phase shift due to the line of electric dipoles is exactly the same result as given in Eq.(7) from the semiclassical analysis.

     In the magnetic case, the electrons pass a solenoid with a magnetic vector potential outside the solenoid 
\begin{equation}
\mathbf{A}_\mu(\mathbf{r})=\hat{\phi}\frac{B_0\sl{A}}{2\pi r}=\frac{(-\mathbf{\hat{i}}y+\mathbf{\hat{j}}x)B_0\sl{A}}{2\pi(x^2+y^2)}=\frac{2\mu(-\mathbf{\hat{i}}y+\mathbf{\hat{j}}x)}{(x^2+y^2)}
\end{equation}
with $\mu=B_0\sl{A}/4\pi$, giving a Hamiltonian for the passing particle
\begin{equation}
H=\frac{1}{2m}\left(p-\frac{e}{c}\mathbf{A}_{\mu}\right)^2
\end{equation}
The shift in the double-slit interference pattern can again be calculated by evaluating the change of phase along the two paths on opposite sides of the solenoid [4]
\begin{equation}
\Delta\phi_{\mu}=\frac{1}{\hbar}\oint \frac{e}{c}\mathbf{A}_{\mu}\cdot d\mathbf{r}=\frac{1}{\hbar}\int dy\left(\frac{e}{c}\mathbf{A}_{\mu y}\right)_{x=+d}-\frac{1}{\hbar}\int dy\left(\frac{e}{c}\mathbf{A}_{\mu y}\right)_{x=-d}=\frac{e B_0 \sl{A}}{c\hbar}
\end{equation}
This phase shift due to the line of magnetic dipoles (solenoid) is exactly the same result as given in Eq.(14) from the semiclassical analysis.

\noindent{\bf DISCUSSION}

     Although the semiclassical analysis given here[8] is straight-forward, the magnetic situation has actually been a subject of considerable controversy[9].  The reasons for the controversy are two-fold.  The first reason involves the interpretation which Aharonov and Bohm[2] gave when they suggested the existence of the magnetic phase shift.  They suggested that the phase shift occurred in the absence of classical electromagnetic forces on the passing electrons.  In other words, there was no classical force involved and so no semiclassical explanation of the phase shift.  Their suggestion implies that the solenoid magnetic flux can be made arbitrarily large and nevertheless the phase shift will never break down due to a finite coherence length for the passing charges.  It also implies that a classical time delay between charges passing on opposite sides of the solenoid can never be observed.  However, the Schroedinger equation itself does not reveal whether classical forces are or are not involved in the magnetic phase shift.  Future experiments may be able to distinguish between the force-dependent semiclassical explanation offered here and the force-free suggestion of Aharonov and Bohm.  In this article we point out simply that there is a natural semiclassical explanation for the presently observed magnetic phase shift.

     The second reason for the controversy arises from misunderstandings of purely classical electromagnetic theory.  Here again there are at least two distinct aspects.  i)Experiments have been performed which attempt to shield the solenoid from the electromagnetic fields of the passing electrons.  The persistence of the phase shift despite the presence of conducting materials shielding the solenoid has been interpreted as excluding the possiblity of a phase shift based upon classical electromagnetic forces.[10]  However, most physicists are unaware that magnetic velocity fields penetrate conducting ohmic materials in a fashion which is completely different from the skin-depth penetration of electromagnetic wave fields.[11]  Thus according to classical electromagnetic theory, the attempts to screen out the electromagnetic fields of the passing charges have actually been ineffectual.  Indeed, it is fascinating to find that within classical electromagnetic theory, the time-integral of the magnetic field due to a passing charge is an invariant independent of any intervening ohmic material.[12]  Furthermore, it is precisely this invariant time-integral which enters the semiclassical explanations of the magnetic phase shift.[13]  ii)In the analysis for the magnetic case, we assumed Newton's third law for the forces between the solenoid and the passing charges.  Only very recently has it been shown within classical electromagnetic theory that the \textit{electric} field of the passing charge causes a $1/c^2$ acceleration-dependent electric field back at the passing charge which has all the correct qualitative features to justify the Newton's third law assumption.[14]  The magnetic phase shift occurs in order $1/c^2$ (since $B_0$ is of order $1/c$), and accordingly the phase shift  requires a relativistic analysis of the classical electromagnetic forces.

\noindent{\bf CONCLUSION}

     Although the physical understanding of the magnetic Aharonov-Bohm phase shift is likely to remain controversial for some time, there is a natural semiclassical explanation for the magnetic phase shift which is exactly parallel to that given for the electric Matteucci-Pozzi phase shift.  Both phase shifts involve lines of dipoles between beams of charged particles---electric dipoles in the one case and magnetic dipoles in the other.  Magnetism has always been more difficult to understand than electrostatics, and this remains true for these phase shifts.

\end{document}